\documentclass{article}

\usepackage{epsfig} 
\usepackage{graphics}
\newcommand{\ket}[1]{\left | #1 \right \rangle}
\newcommand{\bra}[1]{\left \langle #1 \right |}

\begin{document}
\title{
Preparation of spin squeezed atomic states by optical phase shift measurement
}
\author{Isabelle Bouchoule$^1$ and Klaus M{\o}lmer$^2$\\
\small{1: Institut d'Optique, Centre universitaire Bat 503, 
91403 ORSAY Cedex, France}\\\small{ 2: QUANTOP, Institute of Physics and  
Astronomy, University of Aarhus, DK 8000 Aarhus C., Denmark.}}
\maketitle

\begin{abstract}
In this paper we present a state vector analysis of
the generation of atomic spin squeezing by measurement 
of an optical phase shift. 
The frequency resolution is improved when a spin squeezed 
sample is used for spectroscopy in place of an uncorrelated sample. 
When light is transmitted through an atomic sample 
some photons will be scattered out of the incident beam, and this
has a destructive effect on the squeezing. We present quantitative 
studies for three limiting cases: the case of a sample of atoms
of size smaller than the optical wavelength, 
the case of a large dilute sample and the case of a large dense sample.
\end{abstract}
\section{Introduction}

In an atomic sample the population of a state $|a\rangle$ can be
measured non-destructively by a phase shift measurement of an optical field 
which acts on a transition from the state $|a\rangle$.
If the field is not too close to resonance, it
does not drive transitions out of the state $|a\rangle$,
but an initial state vector of the sample with a 
binomial distribution of states with varying populations $n_a$
will be modified, and the quantum mechanical uncertainty of the
number $n_a$ is reduced. This effect has been demonstrated 
experimentally \cite{kuz99,Kuz00}, and further experiments
have shown \cite{juls01}, that two separate atomic ensembles
can be driven into an entangled state by measurements of the total
phaseshifts on an optical field passing through both samples.

A state preparation protocol that applies the outcome of quantum 
non-demolition (QND) measurements
strongly relies on the fact that the measurement entails precisely the
information that is applied in the changes of the state vector.
When light interacts with atoms, apart from a phase shift of
the incident field mode, also scattering out of the field mode
occurs. 
The phase shift results from the interference between the incident 
field and the  component of the scattered field in the incident mode,
whereas scattering out of the incident mode is represented by components 
of the scattered 
wave function orthogonal to the incident wave function 

The scattered photons carry information about the state of the
atoms which is not recorded, and therefore the state of the system
will in general not be the one deduced from the phase shift measurements
alone, but rather an incoherent mixture of the states that one would have
determined if also the scattered photons had been detetected.

The purpose of the present paper is to investigate the importance of
photon scattering for the preparation of atomic states by phase shift
measurements. In particular we shall derive criteria for the
possibility to produce spin squeezed states. 

The paper is organized as follows. 
In Sec. 2, we introduce the concept of spin squeezing and some
useful relations for the mean values and variances of spin operators.
In Sec. 3, we present our model
for phaseshift measurements, and we introduce a formalism that
makes it possible to take photon scattering into account.
In Sec. 4, we analyze the information
given by the registration of phase shifts only. In Sec. 5, we
present simulations and analytical estimates valid for a cloud
which is smaller than the optical wavelength, and for which
photon scattering does not provide any information about
the state of individual atoms. In Sec. 6, we consider the opposite
case where the scattered photons can, in principle, be
traced back to individual atoms in the cloud.  In Sec. 7, we
turn to the more complicated case of a large dense cloud, which
turns out to present the most promising case for 
spin squeezing.  Sec. 8 concludes the paper.

\section{Collective spin representation of an atomic sample}
\label{sec.spinrepresentation}
The term spin squeezing
originates in the treatment of two-level atoms as fictitious spin
$\frac{1}{2}$ particles, $\vec{j}=\frac{1}{2}\vec{\sigma}$, where 
$\vec{\sigma}=(\sigma_x,\sigma_y,\sigma_z)$ are the familiar 
$2\times 2$ Pauli matrices in the basis of atomic states $|a\rangle$ and
$|b\rangle$. For a gas of $N_{at}$ atoms, the collective
spin components
\begin{equation}
\vec{J}=\sum_i \vec{j}_i
\end{equation}
have mean values which characterize the polarization and atomic state
populations of the gas and quantum mechanical uncertainties
which characterize the population statistics. For precision in
spectroscopy and in atomic clocks it is pertinent to have a large
mean spin vector of the sample and to have as small a variance
as possible in a spin component orthogonal to the mean spin.
Assume that the mean spin points in the $x$-direction. Heisenberg's
uncertainty relation states that
\begin{equation}
\Delta J_y \Delta J_z \geq \frac{1}{2}|\langle J_x\rangle |,
\label{Heisenberg}
\end{equation}
For the state with all atoms in their respective
$j_x=\frac{1}{2}$ eigenstate,
the binomial distribution leads to 
uncertainties of
$\Delta J_y = \Delta J_z = \sqrt{N_{at}}/2$, in accord with
the previous inequality.
It was shown by Wineland 
et al \cite{Wine94}, that 
if one can construct spin squeezed states
which do not have the same
uncertainty in the two spin components orthogonal to the mean
spin, one may reduce the variance in a
frequency measurement on $N_{at}$ particles by the factor
\begin{equation}
\xi^2=\frac{N_{at}(\Delta J_z)^2}{\langle J_x\rangle^2}.
\label{Wineland}
\end{equation}

In \cite{Sore01_minvar} states were identified
which for a given $\langle J_x\rangle$ have the smallest possible 
$\Delta J_z$. 
For large $N_{at}$ these are well represented by a Gaussian Ansatz
for the amplitudes on states $|J=\frac{N_{at}}{2},M\rangle$
with different eigenvalues of $J_z$, in 
which case one obtains the approximate relation:
\begin{equation}
\langle J_x\rangle = (J+\frac{1}{2})(1-\frac{2(\Delta J_z)^2}{(2J+1)^2})
\exp(-\frac{1}{8(\Delta J_z)^2}) \sim J\exp(-\frac{1}{8(\Delta J_z)^2}).
\label{gauss}
\end{equation}

Spin squeezed states may be produced in a number of different
ways: by absorption of squeezed light
\cite{Kuzm99PRL}, by controlled collisional interaction in
Bose-Einstein condensates
\cite{Sore01_coll} or in a classical gas \cite{squeezing-Rydberg}, 
by coupling through a single motional degree of freedom 
or through an optical cavity field mode 
\cite{spinsqueezingcav-Luki,single_mode-Klaus}. 
One advantage of the QND scheme, analyzed in the present paper, 
is the automatic matching of
the capability to produce the state and the ability to detect
spin squeezing, which is done by the same kind of measurement.
To verify that the fluctuations in $n_a$ have been reduced, one
has to to show that two subsequent measurements agree (to within
the desired uncertainty). If one can produce a state with reduced 
number fluctuations by means of a QND measurement, one
will also have the resolving power to make use of such reduction
in a high-precision experiment.

\section{A physical setup for phase measurements.}
\label{sec.model}

\begin{figure}
\includegraphics{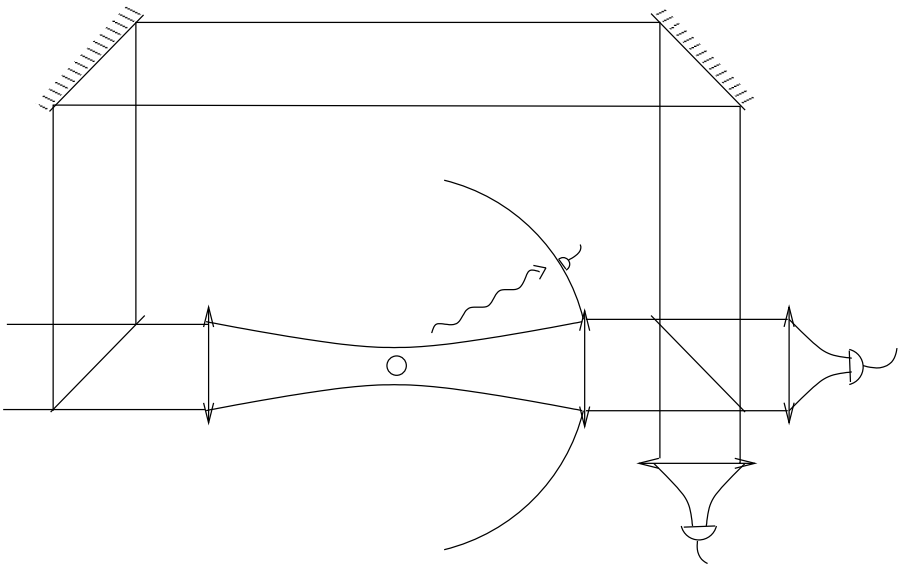}
\vspace{2cm}
\caption{\it Configuration for a spin squeezing experiment. 
Atoms occupying a region in one arm of an interferometer
are illuminated by a component of an optical field, incident 
from the left in the figure. The phase shift of the light field
due to interaction with the atoms in a specific internal state
is registered by the different photocurrents in the two detectors.}
\label{fig.setup}
\end{figure}

In Fig. \ref{fig.setup}, we illustrate a physical setup, where a beam
of light enters a Stern-Gerlach interferometer which contains
an atomic sample in one of its arms. By lenses, the field is focussed
on the sample of transverse dimension $X$.
For simplicity, we assume that the decomposition in plane waves of the
incident photons is uniform for all angles $\theta$ smaller than
the focussing angle $\theta_0$ which, in turn, is so small that
the incident  field is homogeneous across the area of the atomic cloud.
Let $g_0=k \theta_0/(2\sqrt{\pi})$ denote the probability amplitude per unit surface
for a photon to
pass at the center of the mode ($g_0^2dxdy$ gives the probability
that the photon passes in the area $dxdy$ around the center.
A second lens maps the field back onto the inital mode, and the
second beam splitter of the interferometer recombines the optical beams
for read out of the phase shift induced by the presence of atoms
in the lower part of the interferometer. 
The atoms populate states
$|a\rangle$ and $|b\rangle$, and the optical field couples off-resonantly
the state $|a\rangle$ to an auxiliary atomic state, so that a phase shift
on the light field is induced which is proportional to the
population $n_a$.  

We now take into account the scattering of the photons by the atoms.
The scattering is the normal spontaneous emission by the atom, given in the
electric dipole approximation by well known angular distributions for photons
of different polarization. 
Since our purpose is not to  determine the angular
distribution of scattered light, but rather to estimate its damaging 
effect on the
atomic state preparation, we assume simply
that every atom scatters photons isotropically with amplitude $f$.
At low atomic saturation, the scattered photons are coherent with the 
incident field.

We shall present a calculation in which the photons are scattered one at a time
by the sample. 
The scattered part of the wave function of a single photon is entangled 
with the
state of the atomic sample,  since an atomic state with a definite sequence of 
atoms populating the state $|a\rangle$,  
$\ket{\vec{\epsilon}}=\ket{1:\epsilon_1,2:\epsilon_1,... ,N_{\rm at}:\epsilon_{N_{\rm at}}}$
where $\epsilon_i=\ket{a}$ or $\ket{b}$
leads to the photonic wave function
\begin{equation}
\psi_{{\rm scatt}_{\vec{\epsilon}}}=\frac{e^{ikr}}{r}f_{\vec{\epsilon}}
(\Omega)g_0
\end{equation}
where $f_{\vec{\epsilon}}(\Omega)$
depends on the state of the atomic sample. 

In the first Born approximation, $f_{\vec{\epsilon}}(\Omega)$  is
\begin{equation}
f_{\vec{\epsilon}}(\Omega)=
f\sum_{i,\epsilon_i=\ket{a}} e^{i\Delta \vec{k}.\vec{r_i}}\; ,
\label{eq.f}
\end{equation}
where $\Delta \vec{k}$ is the difference between the scattered and the incident
wave vectors.
For $\theta <\theta_0$, $f_{\vec{\epsilon}}(\Omega) \simeq 
f_{\vec{\epsilon}}(0)=f n_a$.

In the Born approximation, the flux of photons is not conserved. 
To remedy this problem, we write the angular part of the photon wave function
far from the atomic sample (sum of the scattered wave function 
and of the incident wave function) in the following  form which is
equivalent to 
the first order Born approximation but which conserves the photon flux.
\renewcommand{\arraystretch}{1.5}
\begin{equation}
f_{\vec{\epsilon}}(\Omega)=\left \{
\begin{array}{l} f_{\vec{\epsilon}}(\Omega)g_0 \mbox{ for }
\theta > \theta_0\\ 
-i\frac{k}{2\pi g_0}\sqrt{(1-\sigma_{scatt})}
\;\; e^{i\frac{2\pi}{k}f n_a |g_0|^2} \mbox{ for} \ \ 
\theta < \theta_0
\end{array} \right . ,
\end{equation} where
\begin{equation}
\sigma_{scatt}=\int_{\theta > \theta_0} \left | f_{\vec{\epsilon}}(\Omega)g_0
\right |^2 d\Omega
\end{equation}
and $f_{\vec{\epsilon}}$ is given by Eq.(\ref{eq.f}).

If the atomic sample  is in one of the states
$\ket{\vec{\epsilon}}$, this state is unchanged by
the transmission of a photon through the interferometer.
The scattering state of the photon, however, depends,
on the argument $\vec{\epsilon}$, and taking into account
the unscattered component in the upper arm, we write this state
in quantum notation as
\begin{equation}
\ket{\psi}_{p,\vec{\epsilon}}=\frac{1}{\sqrt{2}} \left (
\ket{\psi}_{\rm ref}+\ket{\psi}_{\vec{\epsilon}}\right ).
\end{equation}

If the initial state of the cloud is in 
a superposition of different states $\ket{\vec{\epsilon}}$
the joint state of the photon and the atoms becomes the entangled state
\begin{equation}
\ket{\Psi}=
\sum_{\vec{\epsilon}} C_{\vec{\epsilon}}
\ket{\psi}_{p,\vec{\epsilon}}
\ket{\vec{\epsilon}}
=\sum_{\vec{\epsilon}} C_{\vec{\epsilon}}
\frac{1}{\sqrt{2}} \left ( \ket{\psi}_{\rm
ref}+\ket{\psi}_{\vec{\epsilon}}\right )\ket{\vec{\epsilon}}.
\end{equation}

It is at this stage, that the photodetection takes place. The photodetector
1 (resp. 2) of hte interferometer detects photons in the mode 
$\ket{1}=(e^{i\phi}\ket{\psi}_{\rm ref}+ e^{-i\phi}\ket{\psi}_{\rm
inc})/\sqrt{2}$ (resp. $\ket{2}=(e^{i\phi}\ket{\psi}_{\rm ref}- 
e^{-i\phi}\ket{\psi}_{\rm inc})/\sqrt{2}$), with $\ket{\psi}_{\rm inc}$ 
the field transmitted by the lower 
path of the interferometer in the absence of atoms. The detection of
a photon in one of the detectors sketched in Fig.\ref{fig.setup} thus extracts
the corresponding  projection of the state vector  $\ket{\Psi}$. 
This projection causes a non-continuous change of the atomic state vector
amplitudes,
\begin{equation}
\renewcommand{\arraystretch}{1.5}
\begin{array}{l}
C_{\vec{\epsilon}} \rightarrow 
\frac{1}{2} \left ( e^{i\phi}+e^{-i\phi}
\sqrt{(1-\sigma_{scatt})}\;\;
e^{i\frac{2\pi}{k}f n_a |g_0|^2} \right ) C_{\vec{\epsilon}}, \ \ 
\mbox{detection 1,}
\\ 
C_{\vec{\epsilon}} \rightarrow 
\frac{1}{2} \left ( e^{i\phi}-e^{-i\phi}
\sqrt{(1-\sigma_{scatt})}\;\;
e^{i\frac{2\pi}{k}f n_a |g_0|^2} \right ) C_{\vec{\epsilon}},  \ \ 
\mbox{detection 2.}\\
\end{array}
\label{jump12}
\end{equation}
We recall that $n_a$ and $\sigma_{scatt}$ depend on $\vec{\epsilon}$.

The probabilities, $\pi_1$ and $\pi_2$, to detect the photon in detector 
modes 1 and  2 are given by the 
squared norms of the vectors described by the new amplitudes
after application of either projection according to 
(\ref{jump12}).  One may thus simulate the detection process
by chosing one of the two prescriptions, update the amplitudes and 
renormalize the state vector. 
Detection of many photons  is simulated by iterative updating 
of the state vector amplitudes.

In addition to the detection events just described, we have 
identified the possibility for photons to be scattered into 
other directions $\Omega$. The effect of such an event is
of precisely the same character as the projections just described. 
If the photon is detected in the direction  $\Omega$, the corresponding 
projection operator amounts to multiplying each amplitude of the 
initial atomic state with the corresponding scattering amplitude
\begin{equation}
C_{\vec{\epsilon}} \rightarrow f_{\vec{\epsilon}}(\Omega) C_{\vec{\epsilon}},
\ \ \mbox{detection in direction}\ \Omega.
\label{jumpOmega}
\end{equation}
The detector actually has a finite size and detects the photon in
a mode with $f$ centered around $\Omega$ but spread over
$\delta \Omega$. With $\delta \Omega \ll 1/kX$, the scattered wave
function is constant over $\delta \Omega$ and the 
probability to detect a scattered photon within the solid angle 
$\delta \Omega$ in the direction $\Omega$ is thus
\begin{equation}
P(\Omega)=\sum_{\epsilon}
 1/2|g_0|^2|f(\Omega)|^2 |C_{\epsilon}|^2 \delta \Omega
\label{probascatt}
\end{equation}
and the corresponding change of atomic state vector amplitudes
given by (\ref{jumpOmega}).

To simulate the scattering of photons we
divide  the surface of the scattering sphere in sections
by longitudes and lattitudes, and we imagine detectors located
in each section. The  poles of the sphere are in the direction
of the incident beam and the solid angle delimited by 
$\theta < \theta_0$ correponding
to the incident beam 
is of course not covered by  such detectors since  photons emitted 
in this solid angle go in the interferometer. 
For each incident photon, the probability to 
have a click in each detector of the sphere is 
computed.  By adding the probabilities 
we compute $\sigma_{scatt}$ and  we determine
the probabilities to detect the photon in the
detectors 1 or 2. The detector in which the photon
is detected is chosen randomly in the simulation accordingly 
to all the calculated probabilities, and
the state of the atoms is modified according to 
(\ref{jump12}) or (\ref{jumpOmega}).

\section{Information given by the interferometer}
\label{sec.interferometre}

We shall now analyze the states resulting from the interaction with the
field and the detection of the photons.  
In this section we neglect photon scattering,
and we study only the effect of photons measured in detectors 1 and 2. 
We thus assume that $\sigma_{\rm scatt}=0$, in which case we can rewrite the
factors in (\ref{jump12}), and obtain  the state of the atoms
after the detection of $N_1$ photons in 1 and $N_p-N_1$ photons in detector 2,
\begin{equation}
\ket{\psi}=\sum_{\vec{\epsilon}} C_{\vec{\epsilon}}
 \cos\left (\phi-\pi g_0^2\frac{f}{k}n_a\right )^{N_1}
\sin\left (\phi-\pi g_0^2\frac{f}{k}n_a\right )^{N_p-N_1}|\vec{\epsilon}\rangle.
\label{eq.statehom} \end{equation}
In this equation we have ignored a phase factor 
$e^{i\pi fn_a g_0^2 N_p /k}$,  
which corresponds to a phase shift of the state $\ket{a}$ 
or a rotation around $z$ in the spin language.
To avoid such a rotation, one may apply an energy shift on state $|a\rangle$
or an alternative measurement scheme, where atoms in state $|b\rangle$ are
also detected by optical phase shifts.

As a consequence of the photodetections,
the populations of states with a definite number $n_a$ of atoms in 
state $|a\rangle$ are thus multiplied by the factors
\begin{equation} {\cal F}_{N_p}(N_1,n_a)= \cos\left
(\phi-\pi g_0^2\frac{f}{k}n_a\right )^{2N_1} \sin\left (\phi-\pi
g_0^2\frac{f}{k}n_a\right )^{2(N_p-N_1)}.
 \end{equation}

By differentiation with respect to $n_a$, we find that ${\cal
F}_{N_p}(N_1,n_a)$ is peaked at values $n_a^0$ which obey,
\begin{equation}
\tan (\phi+\pi g_0^2\frac{f}{k}n_a^0 )=\pm
\sqrt{\frac{N_p-N_1}{N_1}}.
\label{eq.naointer}
\end{equation}
The values $n_a^0$ correspond of course 
to atomic populations  so that the probability for the photons
to be detected in modes $1$ and $2$ after the interaction
are in agreement with the ratios $N_1/N_p$ and $N_2/N_p$
observed by the measurement.

To estimate the width of ${\cal F}_{N_p}(N_1,n_a)$, we calculate the second derivative
of ${\cal F}_{N_p}(N_1,n_a)$ at $n_{a}^0$. We find that
\begin{equation}
\frac{\partial ^2}{\partial {n_a}^2}{\cal F}_{N_p}(N_1,n_a) (n_a^0)= -4 N_p \left (
\pi g_0^2\frac{f}{k}\right )^2 {\cal F}_{N_p}(N_1,n_a^0).
\label{eq.width}
\end{equation}
So, the more photons that are transmitted, the narrower is 
the width of the 
peaks in ${\cal F}_{N_p}(N_1,n_a)$. The width
does not depend on the initial relative phase $\phi$
between the two arms of the interferometer or on the result of the
measurement.
 If we suppose that  ${\cal F}$ is gaussian, the following equation
gives the rms width of ${\cal F}$ in $n_a$
\begin{equation}
\Delta n_{a_{\rm int}}=\frac{1}{2 \pi g_o^2 \frac{f}{k}\sqrt{N_p}}
\label{eq.Dnaint}
\end{equation} 

The equation (\ref{eq.naointer}) has several solutions due to the 
two possible signs, and due to the periodicity of the $tan$-funtion.
If several such solutions lie within the initial binomial
distribution of $n_a$, the state obtained after the measurement will be a 
coherent superposition of spin states with different mean values of $J_z$,
i.e., a kind of ``Schr\"odinger cat".
However, a change of $N_1$ or $N_p-N_1$ by unity changes the relative 
phase between peaks by $\pi$.
Thus, with realistic photon detectors with an efficiency smaller
than unity, the relative phase is unknown and the system is
described by a statistical mixture of the states.

In order to obtain spin squeezing, we want to ensure that only a single
value of $n_a^0$ inside the initial distribution 
obeys Eq.(\ref{eq.naointer}), so that the 
detection unambiguosly leads to a more narrow distribution in
$n_a$. This requires
\begin{equation}
\frac{\pi}{2\pi g_0^2 f/k} >  \sqrt{N}
\; \mbox{ and } \; \frac{\Phi k}{\pi g_0^2 f} \gg \sqrt{N}
\label{eq.pasdechat}
\end{equation}

Eq.(\ref{Wineland}) shows that it is not enough to reduce
the uncertainty in $J_z$ to have useful spin squeezing, one must
also ensure that the mean value of $J_x$ remains large. The outcome of
the interferometric detection
is close to ideal in this respect. The resulting state vector
has amplitudes on the different $J_z$ eigenstates which follow a
Gaussian distribution very well, and the approximation (\ref{gauss}) for the
mean spin is close to the maximum possible value for any given
variance of $J_z$.

Using Eqs.(4,18) we obtain:
\begin{equation}
\xi =\frac{1}{\sqrt{N_{at}}\sqrt{N_p}\pi g_0^2\frac{f}{k}}
e^{\pi^2 g_0^4f^2N_p/2k^2}
\end{equation}
The minimum value of $\xi$, for a fixed $N_{at}$, is 
\begin{equation}
\xi_{\rm Min}=\sqrt{\frac{1}{N_{at}}}.
\label{xiperf}
\end{equation}
and is obtained for the photon number $N_p=k^2/\pi^2 g_0^4 f^2$. 
This value of $\xi$ is the minimum value allowed as shown in \cite{Wine94}.

For a large number of atoms, the squeezing factor (\ref{xiperf})
can be really significant.
The production of spin squeezed states by QND detection is susceptible, 
however, to two possible drawbacks caused by scattering of photons:

\begin{itemize}
\item{scattered photons carry information about $J_z$, so that this 
quantity could in principle be known better than the width of 
${\cal F}_{N_p}(N_1,n_a)$,
and according to Eq.(\ref{gauss}), the mean spin will be reduced}

\item{ scattered photons carry information about 
the spatial distribution of atoms in the state $\ket{a}$.
The state is then  no longer symmetric under exchange of the particles,
$J$-values smaller than $N_{at}/2$ become populated, and the
mean spin is accordingly reduced.}
\end{itemize}

We shall now turn to a quantitative analysis of these effects.

\section{Small cloud}
\label{small.cloud}

In this section we consider the case of a cloud of atoms
confined to a region in space smaller than the optical
wavelength. This implies that the scattered photons
will not contain information about the individual
atoms in the ensemble. They will, however, carry information
about $n_a$ that is not known to the experimentalist
who measures only the fields by the detectors 1 and 2. 

\subsection{Numerical simulations}
\label{num.sim}

In a numerical simulation of the detection process we place atoms
randomly in space according
to a gaussian probability distribution, 
and we assume an initial state where all atoms are in
$(\ket{a}+\ket{b})/\sqrt{2}$. No restriction is made on the state at later
times, which is expanded on the whole space of dimension $2^{N_{at}}$.

\begin{figure}
\centerline{\includegraphics{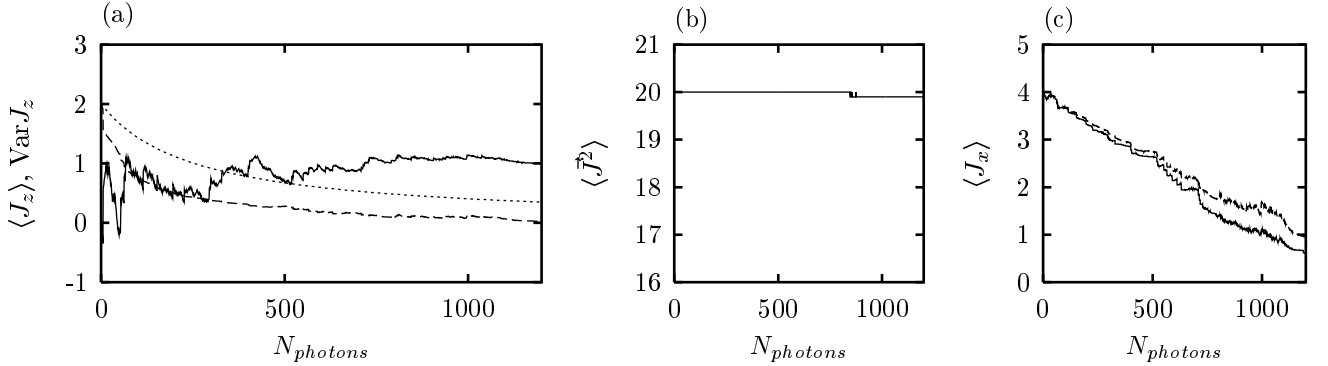}}
\caption{\it
Results of a single simulation with 8 atoms whose spatial positions follow
 a gaussian law with a width of $10^{-2}\lambda /2\pi$. The angular
 spread of the incident beam is $\theta_0=0.45\pi$, close to
 the best that can be achieved.
(a) Solid line : $\langle J_z \rangle$ as a function of the number of
photons launched on the atoms.
Long-dashed line : $\langle (J_z -\langle J_z \rangle)^2\rangle$.
Short-dashed : Expected evolution of
$\langle (J_z -\langle J_z \rangle)^2\rangle$ according to
Eq.(\ref{eq.Dnaint}), taking into account the width of the initial
distribution.
(b) Evolution of $\langle \vec{J}^2 \rangle$. The
slight
decrease shows that the state vector acquires only small components
outside the symmetric subspace. 
(c) Evolution of the mean value of the spin in the horizontal
direction (upper trace) and of the component of the spin
along the expected direction (lower trace).
}
\label{fig.1histpres}
\end{figure}

Fig.\ref{fig.1histpres} presents the evolution obtained for
one particular history for a cloud of 8 atoms confined to 
a spatial region of dimension $\lambda/100$
and interrogated by a beam which is focussed on the atoms with an 
angular aperture of $\theta_0=0.45\pi$.
The variance of the  distribution of atoms in $\ket{a}$
({\it i.e.}, $\langle (J_z -\langle J_z \rangle)^2\rangle$) is 
plotted as well
as the value expected from the results of section \ref{sec.interferometre} 
({\it i.e.}, the multiplication of the 
initial distribution with the function ${\cal F}_{N_p}(N_1,n_a)$).

The value of $\langle \vec{J}^2 \rangle$ keeps almost the initial maximal
value of $20$ which indicates that the cloud stays in a symmetric
state. This is expected as the atoms are closer to each other than
$\lambda$ and then it is not possible to discriminate between the
atoms with the scattered photons.

The last graph plots the value of  $\langle J_x\rangle $ and of 
$\sqrt{\langle J_x\rangle^2 +\langle J_y \rangle ^2}$
which is the length of the mean spin in the horizontal plane,
and which may be larger than the $x$-component because of small angular
spin rotations that occur during photon scattering.

\begin{figure}
\centerline{\includegraphics{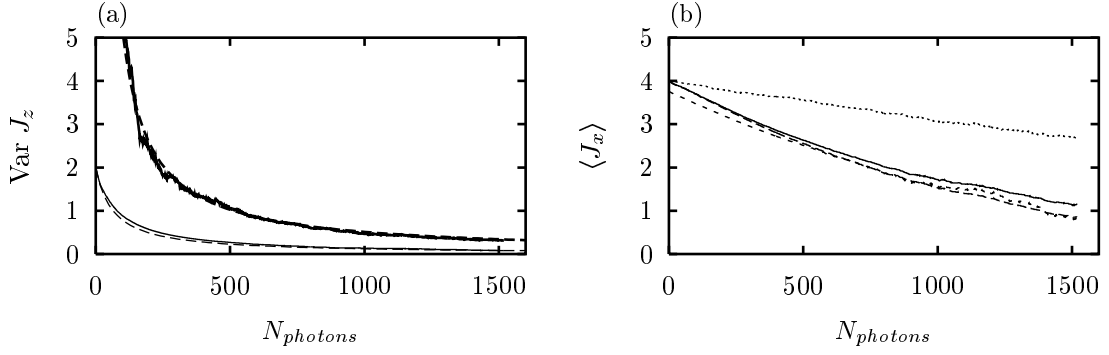}}
\caption{\it 
Average over 90 histories.
(a) Thin solid line: evolution of
$\langle (J_z -\langle J_z \rangle)^2\rangle$ as a function of the number of photons launched on
the
atoms. Note that this quantity is not measurable as $\langle J_z
\rangle$ depends on the particular history.
The dashed thin line gives the value expected from the
scattering process only. It is a convolution between the initial distribution
and a distribution of width given by Eq.(\ref{eq.sqdiff}), both assumed
to be gaussian.
Solid fat line: evolution of
$\langle (J_z-J_{z_{calc}})^2\rangle$ where $J_{z_{calc}}$ is
deduced, for each history, from Eq.(\ref{eq.naointer}) and from the knowledge
of the number $N_1$ of photons detected in 1 and
the number $N_2$ of photons
detected in 2. This quantity agrees with
Eq.(\ref{eq.Dnaint}) shown by the dashed fat line.
(b) Solid line and short dashed line : Evolution of 
$\sqrt{\langle J_x\rangle^2 +\langle J_y \rangle ^2}$ and 
$\langle J_x\rangle $. A rotation of
$\pi/k f g_0^2 N_{photons}$ is applied in the $xy$-plane to 
compensate for the light shift.
Dashed line : expected behavior due to the squeezing realized  by
scattering.
Dotted line : expected behavior if the
width of the $J_z$ distribution were only given by the interference 
detection.
 }
 \label{fig.numpres_2}
 \end{figure}

Figure \ref{fig.numpres_2} shows the variance of $J_z$ and the mean value of $J_x$ and of
the largest projection of the spin orthogonal to the $z$-axis.
These results are obtained as the average over 90 independent realizations of
our simulation. 
Both in Fig.\ref{fig.1histpres}., and in Fig.\ref{fig.numpres_2},
we observe that the actual width in $J_z$ is smaller than the
one concluded from the interferometric measurement~:
the fact that the atoms are coupled to other modes of
the field also leads to  squeezing. We recall, however, that the quantity
$\langle (J_z -\langle J_z \rangle)^2\rangle$ is not a measurable
quantity as $\langle J_z\rangle $ depends on the particular history.
To exploit the squeezing due to scattering, one has to deduce
$\langle J_z\rangle $ for each experiment by keeping track of the
scattered photons. 

\subsection{Analytical estimates}

The effect of the scattered photons can be computed analytically.
Taking $\vec{r}_i=\vec{0}$ for all $i$, Eq.(\ref{eq.f}) and
Eq.(\ref{jumpOmega}) show 
that the atomic state vector amplitudes are multiplied
simply by the coefficient $\sqrt{\sigma_1} n_a$ after the detection of 
a scattered photon, and by the coefficient $\sqrt{1-\sigma_1}n_a$ in
the absense of scattering where $\sigma_1=f^2g_0^22\pi(1+\cos\theta_0)$ is
the scattering probability per atom in state $|a\rangle$.

After the detection of $N_{\rm scatt}$ out of  a total number
of $N_p$ photons, the wave function of the atoms becomes
\begin{equation}
\ket{\psi}=\sum_{n_a} C_{n_a}
 \left ( \sqrt{\sigma_1} n_a \right )^{N_{\rm scatt}}
 \sqrt{1-\sigma_1 n_a^2}^{N_p-N_{\rm scatt}}|\vec{\epsilon}\rangle,
\label{eq.statediff}
\end{equation}
so the probability distribution for $n_a$ is multiplied by
\begin{equation}
{\cal G}(N_{\rm scatt},na)=(\sigma_1 n_a^2)^{N_{\rm scatt}}
 \left (1-\sigma_1 n_a^2 \right )^{N_p-N_{\rm scatt}}
\end{equation}

This function is maximum on
\begin{equation}
n_{a_{0_{\rm scat}}}=\sqrt{\frac{N_{\rm scatt}}{
2\pi(1+\cos\theta_0)g_0^2f^2 N_p}}
\label{eq.naodiff}
\end{equation}
and its width can be estimated by assuming a gaussian
shape with the same second derivative
at the peak value 
\begin{equation}
\Delta {n_a}_{\rm scat}
\simeq\frac{1}{\sqrt{8\pi(1+\cos\theta_0)g_0^2f^2 N_p}}.
\label{eq.sqdiff}
\end{equation}
This width is always smaller than the width in $n_a$ of the function
$\cal{F}$ because
$g_0=k\theta_0/(2\sqrt{\pi})$, which is about the inverse of the transverse size  of the
beam on the cloud is always smaller than $k$,
\begin{equation}
\frac{\Delta {n_a}_{\rm scat}}{\Delta {n_a}_{\rm
int}}=\frac{1}{4}\frac{g_0}{k}\frac{1}{1+\cos(\theta_0)} < 1
\end{equation}

In every single realization, the width of the $n_a$ distribution
is thus set by the number of scattered photons. After averaging
over the unkown number of scattered photons we recover the
broader distribution determined by the interferometer readout
$N_1$ and $N_2$.
Since the scattered photons do not drive atoms out of or into
the state $|a\rangle$,
the atomic density matrix obtained by an average over the
number of scattered photons has the same diagonal elements
in the basis $\{\ket{J,M}\}$ as the pure state that one
would expect without photon scattering. 
But, the coherence terms of the density matrix  are different
from that of the pure state, and therefore the length of the
mean spin will be altered by the scattering events.
This is seen in Fig.\ref{fig.numpres_2}  b), where the three lower curves show
the actual value of $\langle J_x\rangle$, of $\sqrt{\langle J_x\rangle^2
+\langle J_y\rangle^2}$ and of the estimate (\ref{gauss}), based
on the small variance of $J_z$ due to the scattering. 
There is excellent agreement between these curves. The upper curve
shows the larger value of the mean spin, that one would have obtained in
the absence of scattering.

Our numerical simulations and our analytical estimates show that
the effect of the scattering is to produce stronger
squeezing than the interference measurement.
After averaging over the unresolved scattering histories
this squeezing does not affect the $n_a$ populations. But, its effect
is to reduce the mean spin $\langle J_x\rangle$.
In principle one could determine the number of scattered photons
by the difference betwen the number of incident photons and the number
of photons detected in the interferometer. For a poissonian source
of light, however, this number cannot be determined to a higher precision
than $\sqrt{N_p}$, which turns out to be larger than the
required precision on the loss in photon number due to scattering.

As pointed out in section \ref{sec.spinrepresentation}, the
pertinent factor for spin squeezing is $\xi$ defined in Eq.(\ref{Wineland}).
Using Eq. (\ref{eq.Dnaint}) for $\Delta J_z$ and  Eqs.(4,25) to 
determine $\langle J_x \rangle$, we obtain

\begin{equation}
\xi =\frac{1}{\sqrt{N_{at}}\sqrt{N_p}\pi g_0^2\frac{f}{k}}
e^{2\pi g_0^2f^2N_p}
\end{equation}
The minimum value of $\xi$, for a fixed $N_{at}$, is 
\begin{equation}
\xi_{\rm Min}=\sqrt{\frac{4e}{\pi N_{at}}}\frac{k}{g_0}.
\end{equation}
and is obtained for the photon number $N_p=1/(\sqrt{4\pi} g_0^2 f^2)$. 
Because the size of the incident beam is larger than the wave length, $g_0<k$
and $\xi_{\rm Min}$ is larger than in the ideal case (\ref{xiperf}). 
 
\section{Large dilute cloud.}
\label{sec.numloin}

The effect of the scattering in the case of a cloud of extension
larger than $\lambda$ is very different from the case of a small
cloud. The number of scattered photons will still, as in the
case of a small cloud, give us information on the total number
of atoms in $\ket{a}$ and thus scattering will by itself produce
squeezing.
But in the case of a big cloud the angular distribution
of the scattered photons gives also information on the position of the atoms
in $\ket{a}$ and the atomic system therefore looses its invariance under permutation
of the particles. The system will thus no longer be
fully represented by the ($N_{at}+1$) Dicke states of
maximal $J=N_{at}/2$.
This does not affect the $z$ component of the spin but it
strongly reduces the horizontal component of the spin.

\subsection{Numerical simulation}
\label{num.sim.6}

\begin{figure}
\centerline{\includegraphics{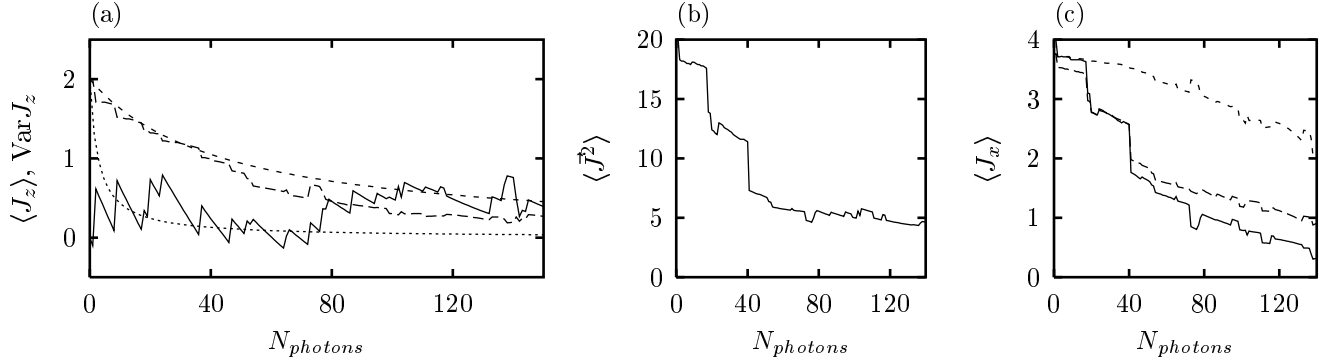}}
\caption{\it
The cloud contains 8 atoms spatially distributed according to a 3D gaussian
probability law of rms width 
$10\lambda/2\pi$. The incident beam, focussed
on the atoms, has an angular spread
$\theta_0=2\sqrt{\pi}g_0/k=2\sqrt{\pi}*0.05=0.14$.
(a) : Evolution of $\langle J_z \rangle$ (solid line) and
$\langle (J_z-\langle J_z \rangle)^2\rangle$ (long dashed line) as
functions of the number of photons which are launched on the atoms.
The dotted line indicates the value of  
$\langle (J_z-\langle J_z \rangle)^2\rangle$
expected for a dense cloud of 8 atoms illuminated by the same beam (Eq.
(\ref{eq.sqdiff})).
The short dashed line gives the result of Eq.(\ref{eq.squeezescatt_loin}).
(b) : Evolution of $\langle \vec{J}^2 \rangle$. The first drop from the
maximal value 20 occurs at the first detection of a scattered photon.
(c) : Evolution of the mean horizontal spin (solid line). In dashed
lines is shown the expected value of $\langle J_x \rangle$ according
to Eq.(\ref{gauss}) for  $J=N_{at}/2$.
The result of Eq. (\ref{eq.JxvsJznonSym}) is shown with the short-dashed
line.
}
\label{fig.atomesloinscatt}
\end{figure}

In order to analyse the effect of the scattering alone, we have performed
simulations of the evolution of the atomic state under detection
of scattered photons. The directions of photodetection were
chosen according to the procedure outlined in Sec. III,  and
Fig. \ref{fig.atomesloinscatt} presents the result of a single
simulation. In this simulation, we see a decrease of
$\langle (J_z-\langle J_z \rangle)^2\rangle$ as well as a
departure from the symmetric space (decrease of $\langle \vec{J}^2 \rangle$).
It is also clear that the decrease of
$\langle (J_z-\langle J_z \rangle)^2\rangle$
is  not as rapid as suggested by Eq.(\ref{eq.sqdiff})
which was valid for the small cloud.

Let us present a naive argument for the decrease
of $\langle (J_z-\langle J_z \rangle)^2\rangle$:
Assume that the cloud is in a state $\ket{\vec{\epsilon}}$
where each atom is either in $\ket{a}$ or $\ket{b}$. This state is 
unaffected by the interaction with the photon field.
Let $p$ denote the probability that a photon is scattered by the cloud
of atoms.
In the absence of superradiance
we expect $p=\sigma_1 n_a$.
The number of scattered photons 
$N_{scatt}$ thus has mean value $p N_p$ and uncertainty 
$\sqrt{p N_p}$, and we estimate
$n_a$ by $N_{scatt}/\sigma N_p$ with an uncertainty of 
$\Delta n_a=\sqrt{n_a/\sigma N_p}$.
If the state of the cloud expands over different $n_a$, the measurement
of $N_{scatt}$ will reduce the width of the distribution by multiplying
with a function of width $\Delta n_a=\sqrt{n_a/\sigma N_p}$.
Taking $\sigma=4\pi g_0^2 f^2$ 
and an initial state where all atoms are in $(\ket{a}+\ket{b})/\sqrt{2}$ 
\begin{equation}
\langle (J_z-\langle J_z \rangle)^2\rangle=
\frac{1}{\frac{4}{N_{at}}+  \frac{8\pi g_0^2 f^2 N_p}{N_{at}}}
\label{eq.squeezescatt_loin}
\end{equation}
This function is plotted in Fig.\ref{fig.atomesloinscatt} (a).
It reproduces quite well the
numerical evolution, although the numerical evolution shows a faster squeezing.

After averaging over histories, as discussed in section 
\ref{small.cloud},
the squeezing due to scattering has no effect on the distribution of $n_a$
but it will reduce the horizontal component of the total spin.
The upper dashed line in figure
\ref{fig.atomesloinscatt}(c) predicts the reduction due to this 
effect, according to Eq.(\ref{gauss}) taking $J=N_{at}/2$.
We see that the horizontal spin determined in our simulation is even 
smaller than this estimate. The reason for
this is that the state of the cloud is not in the symmetric subspace as
assumed when we put $J=N_{at}/2$ in Eq.(\ref{gauss}).
The state of the cloud may be expanded on subspaces
with different total $J$, and the mean value of $J_x$ is averaged over these
different subspace components. 
To check the consistency of this picture we compare the decrease of
$\langle \vec{J}^2\rangle$ with that of $\langle J_x\rangle$.
The maximum $\langle J \rangle$ for a given$\langle \vec{J}^2\rangle=\langle J(J+1)\rangle$
is $\sqrt{1+4\langle J(J+1)\rangle}/2-1/2$ and
it is obtained if only the subspace with  $J=\langle J \rangle$ is populated.
If we assume that the $J_z$ distribution (centered around zero) 
is the same in all subspaces we estimate that
\begin{equation}
\langle J_x\rangle \leq e^{-\frac{1}{8\mbox{Var} J_z}}
\left ( -\frac{1}{2}+\frac{\sqrt{1+4\langle J(J+1)\rangle }}{2}\right )
\label{eq.JxvsJznonSym}
\end{equation}
The curve corresponding to the right-hand side of this inequality is
plotted in Fig.\ref{fig.atomesloinscatt}.
 It reproduces 
quite well the decrease of $\langle J_x \rangle$.

\subsection{Analytical estimates}
\label{sec.atomespres}

If the atomic cloud is dilute enough and not too large
it is possible to know by which atom any photon  has been emitted. 
In other words, it
is possible  to design an optical system which collects all the photons emitted
outside the incident mode and which produces a separate image of each atom.
This implies that the flux of scattered photons is only proportional to 
$n_a$ and not to $n_a^2$: there is no superradiance. 
Then the scattered photons give information on the total number of
atoms in $\ket{a}$ but also on which atoms are in $\ket{a}$.
The second effect damages the squeezing realised by the interference
measurement as it decreases correlations between atoms.
 When a photon is transmitted, the probability that it is scattered by an atom
in $\ket{a}$ is approximately $4\pi g_0^2 f^2$.
Thus after about $N_p=1/4\pi g_0^2 f^2$ photons
have been transmitted, we know almost with certainty if the atom is
in $\ket{b}$ or $\ket{a}$, and there is then
almost no correlation between the atoms and hence no squeezing. 
Indeed, as each atom
is either in $\ket{a}$ or $\ket{b}$, $\langle j_x\rangle=0$ for each atom.
In order to observe squeezing the number of photons used in the 
experiment is limited and this in turn limits the reduction in
Var$(J_z)$.  

Within a more quantitative analysis of the effect of 
scattering let $\phi$ denote the flux of incident photons.
As long as an atom scatters no photons, its state
evolves according to the effective Hamiltonian
\begin{equation}
H_{nh}=-\frac{i}{2}\phi 4\pi g_0^2f^2\ket{a}\bra{a}.
\end{equation}
The probability to scatter a photon during $\delta t$ is
\begin{equation}
P\delta t=\bra{a}\psi\ket{a}^2\phi 4\pi g_0^2f^2 \delta t.
\end{equation}
After averaging over histories only
the coherence term $\sigma_{ab}$ of the density matrix  evolves 
and it obeys
\begin{equation}
\frac{d \sigma_{ab}}{dt}=-\frac{\phi}{2}4\pi g_0^2f^2 \sigma_{ab}.
\end{equation}
Thus, $\langle j_x \rangle =1/2( \sigma_{ab}+\sigma_{ab}^*)$ follows
the same exponential decay, and it follows that the total spin 
$J_x=\sum j_{x_i}$ decreases as
\begin{equation}
\langle J_x\rangle_t=\langle J_x\rangle_0e^{-2\phi \pi g_0^2f^2 t}
=\langle J_x\rangle_0e^{-2 N_p \pi g_0^2f^2 }
\end{equation}

The reduction of $\langle J_x\rangle$ due to the interferometric
dectection, estimated using Eq.(\ref{gauss}) is less  important than 
the reduction of $\langle J_x\rangle$ due to scattering and 
we will neglect it here.
Var$(J_z)$ is given by Eq.(\ref{eq.Dnaint}), and 
the squeezing factor writes
\begin{equation}
\xi=\frac{1}{\sqrt{N N_p}\pi g_0^2 \frac{f}{k}} 
e^{2 N_p \pi g_0^2 f^2}.
\end{equation} 
Its minimum value is
\begin{equation}
\xi_{\rm Min}=\sqrt{\frac{2e}{\pi N}}\frac{k}{g_0},
\label{eq.bestxidilute}
\end{equation}
and it is obtained for $N_p=1/(4\pi g_0^2 f^2)$.

Although the physical regimes are very different, the result is similar
to Eq.(26).

\section{Large dense cloud}
We now turn to an analysis of the situation of many atoms in a large
cloud with a  density which is large compared to
$1/\lambda^3$. In this case, we are not able to carry out 
simulations, and we are
therefore restricted to an analytical approach. 
We shall make the assumption of a dense 
cloud, so that it may be divided into a large number of 
cells of size $a^3$ smaller than $\lambda^3$
but still containing a large number of atoms.
For simplicity, we assume that the distribution of atoms is uniform
over a box of size $L_x=L_y,L_z$.
As
the size of the cell is smaller than $\lambda^3$, the scattering of
photons does not bring information on the repartitioning of the atoms in
$\ket{a}$ inside each cell and the state of the cloud
can be described by states that are symmetric under exchange of particles 
inside each cell.
The states $\ket{\vec{n}}=\ket{n_1,n_2,...,n_M}$ with well defined
number $n_1$, $n_2$,...,$n_M$ of atoms in $\ket{a}$ in the cells
1,2,...,M are not modified by the scattering. 
The number of atoms per cell is sufficiently large so that
we consider that the number of atoms in $\ket{a}$, on the order of 
$N_{cell}/2$, can be considered as a continuous variable with 
fluctuations of order $\sqrt{N_{cell}}/2$.

\subsection{Initial state of the cloud}
Initially, all the atoms are in the state
$1/\sqrt{2}(\ket{a}+\ket{b})$.
The expansion of the state of the cloud  on the
basis $\{\ket{\vec{n}}\}$ is then
\begin{equation}
\ket{\psi}_{at}=\int dn_1 \int dn_2 ...\int d n_M
c(n_1)c(n_2)...c(n_M) \ket{n_1,n_2,...,n_M}
\end{equation}
where $c(n)$ is the square root of the binomial distribution of mean 
value $N_{cell}/2$
which we will approximate by a gaussian distribution.

We will now define the  new variables
\begin{equation}
\left \{
\begin{array}{l}
N_{c_{i_x,i_y,i_z}}=\sqrt{\frac{2}{M}}\sum_{\vec{l}} n_{\vec{l}}\cos(\vec{K}_{\vec{i}}.\vec{r}_{\vec{l}})\\
N_{s_{i_x,i_y,i_z}}=\sqrt{\frac{2}{M}}\sum_{\vec{l}}
n_{\vec{l}}\sin(\vec{K}_{\vec{i}}.\vec{r}_{\vec{l}})\\
N_0=\sqrt{\frac{1}{M}}\sum_{\vec{l}} n_{\vec{l}}
\end{array}
\right .
\label{eq.defNcs}
\end{equation}
with 
\begin{equation}
\left \{
\begin{array}{l}
\vec{K}_{\vec{i}}=\frac{2\pi}{aN_x}i_x \vec{x}^0 +
\frac{2\pi}{aN_y}i_y \vec{y}^0+
\frac{2\pi}{aN_z}i_z \vec{z}^0\\
\vec{r}_{\vec{l}}=a l_x\vec{x}^0 +
a l_y\vec{y}^0 +
a l_z\vec{z}^0
\end{array}
\right .
\end{equation}
where $i_z<0$ or ($i_z=0$ and $i_x<0$) or ($i_z=i_x=0$ and $i_y<0$).

Note that all the operators $N_\alpha$ and
$N_\beta$ commute since they
are all diagonal in the basis $\{\ket{\vec{n}}\}$.

The initial state of the cloud can be expanded on the new basis of
eigenstates
\begin{equation}
\ket{\psi}_{at}=
\begin{array}[t]{l}
\int dN_0 \int dN_{c_1}\int d N_{s_1} ...\int d
N_{c_{(M-1)/2}} \int d N_{s_{(M-1)/2}}\\
C(N_0)h(N_{c_1})h(N_{s_1})... h(N_{c_{(M-1)/2}})h(N_{s_{(M-1)/2}})
\ket{N_0,N_{c_1},N_{s_1},...,N_{c_{(M-1)/2}},N_{s_{(M-1)/2}}}
\end{array}
\end{equation}
where $h$ is a gaussian centered on 0 with an rms width of 
$\sqrt{N_{cell}/2}$ 
($h^2$ has a width $\sqrt{N_{cell}}/2$ equal to the width of $c^2(n_i)$) and
$C$ has the same width but is centered on $\sqrt{M} N_{cell}/2$.
Initially, there is no correlation between the distributions 
in $N_{{s/c}_i}$ and they are all the same.

The vector space can be seen as a tensor product of (M+1)/2 spaces:
The space acted upon by $N_0$, and for each $i$, the space
${\cal E}_i$ acted upon by  the operators $N_{c_i}$ and $N_{s_i}$.  
All these spaces
have infinite dimension and they admit as basis states
$\ket{N_{c_i},N_{s_i}}$ where $N_{c_i},N_{s_i}$ are real.
This basis will turn out to be useful for the description of the
state vector dynamics due to photon scattering.

\begin{figure}[ht]
\includegraphics{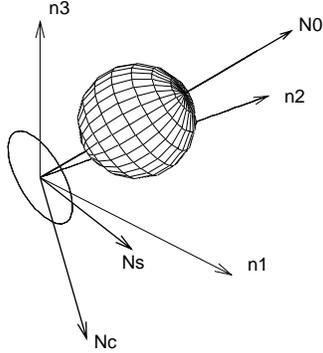}
\caption{\it Inital distribution among the states
($\ket{n_1,n_2,...,n_M}$) or the states
($\ket{N_0,N_{c_1},N_{s_1},...,N_{c_{(M-1)/2}},N_{s_{(M-1)/2}}}$)
in the case where there are only three cells ($M=3$).
Represented is a surface on which the population is constant.
The circle surrounding the origin is the projection of the
sphere on the $(N_c,N_s)$ plane.}
\end{figure}

\subsection{Effect of the scattering}
Due to energy and momentum conservation in the scattering process,
a fluctuation in the cloud will couple to the light only 
if its Fourier transform  has components $\vec{K}_i$ on the
scattering sphere defined as the vectors  $\vec{k}_d$ which fulfill
$|\vec{k_d}-\vec{k}_{inc}|=k$. 

\begin{figure}
\centerline{\includegraphics{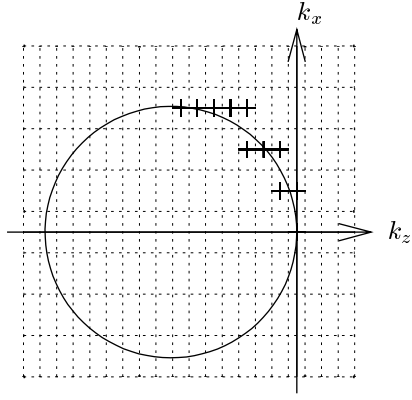}}
\caption{\it Diffusion sphere and components of the
discret Fourrier decomposition which participate to the
diffusion}
\label{fig.Spherediff}
\end{figure}

We will now assume for simplicity that the Fourier transform
associated with each $\vec{K}_i$ 
are uniform over disjoint volumes $\delta k_x=2\pi/L_x$, $\delta k_y=2\pi/L_y$
and $\delta k_z=2\pi/L_z$, shown as small rectangles in
Fig. \ref{fig.Spherediff}. The circle in Fig. \ref{fig.Spherediff}
depicts the scattering sphere and crosses represent the dicrete 
$\vec{K}_i$ wave vectors which contribute to the scattering.
With this approximation, the observation of a scattered photon
gives information on the modulation of the atomic population in state
$|a\rangle$ with the corresponding discrete wave vector.

Let us consider a $\vec{K}_i$ which is on the scattering
sphere and the associated solid angle $\delta \Omega_i$.
The effect on the state of the system associated with the 
detection of a photon in this solid angle is given by the
operator
\begin{equation}
{\cal P}(\Omega) = \sqrt{\delta \Omega_i}g_0 f \sum_{\vec{l}}
n_{\vec{l}}
e^{i(\vec{k_d}-\vec{k}_{inc})\vec{r}_{\vec{l}}}=
\sqrt{\delta \Omega_i}
g_0 f \sqrt{\frac{M}{2}} (N_{c_{\vec{i}}}+iN_{s_{\vec{i}}}).
\end{equation}
This operator changes the relative phase 
between the component with
fluctuations in cosinus and sinus components. 
Because the operator in Eq.(40) acts only on the space ${\cal E}_i$,
no correlations between fluctuations at different wave vector
appear 
and the state of the cloud will thus stay on the form
\begin{equation}
\ket{\psi_0}\ket{\psi_1}...\ket{\psi_{(M-1)/2}}
\label{eq.productstate}
\end{equation}
where $\ket{\psi_i}$ is a state of the space ${\cal E}_i$.
 Note that this result is an approximation which relies on our assumption
that the Fourier transform of the fluctuations at different $\vec{K}_i$ 
are disjoint.
 Furthermore, the assumption that atoms are uniformly distributed over 
our spatial grid is important. 
Indeed, if the number of atoms
in $\ket{a}$ in different cells is not the same for each cell, 
the initial state would present correlations  
between $K_i$'s and knowledge about the state in the subspace ${\cal E}_i$
would also implies knowledge about the state in other ${\cal E}_j$.  

If a photon is detected in the incident mode, 
the effect on the state of the atoms, to second order in $f$, is
given by
\begin{equation}
{\cal P} =\prod_i
\sqrt{1- g_0^2 f^2 \frac{M}{2}\delta \Omega_i
\left ( N_{c_i}^2+N_{s_i}^2
\right ) }.
\end{equation}
Thus, in this approximation, neither scattering nor absence of
scattering yields correlations between different $K_i$'s, 
and the state of the cloud will stay in a product state
as in Eq.(\ref{eq.productstate}).

After transmission of $N_p$ photons and  the detection of $N_{sc}$ 
photons scattered in the direction
$\vec{k}_{diff}=\vec{k}_{inc}-\vec{K}_i$,
the wave function in the space ${\cal E}_i$ becomes
\begin{equation}
\ket{\psi_i}=
\int dN_{c_i}\int d N_{s_i}
\left ( N_{c_i} + i N_{s_i} \right )^{N_{sc}}
\left ( \sqrt{1-g_0 ^2 f^ 2 \frac{M}{2} \delta \Omega_i\left ( N_{c_i}^2 +N_{s_i}^2 \right )}\right )^{N_p-N_{sc}}
h(N_{c_i})h(N_{s_i})\ket{N_{c_i},N_{s_i}}
\end{equation}
Thus, the population of the states $\ket{N_{c_i},N_{s_i}}$ is
multiplied by the factor
\begin{equation}
{\cal G}(\sqrt{N_{c_i}^2+N_{s_i}^2})
\begin {array}{l}
=\left ( N_{c_i}^2 +  N_{s_i}^2 \right )^{N_{sc}}
\left ( 1-g_0 ^2 f^ 2 \frac{M}{2} \delta \Omega_i\left (
N_{c_i}^2 +N_{s_i}^2\right) \right )^{N_p-N_{sc}},
\end{array}
\end{equation}
which depends only on $N_{c_i}^2 +  N_{s_i}^2$. This is expected
as the detection of the scattered photons does not reveal any information
about the phase of the spatial grating in the cloud 
of atoms in $\ket{a}$.

A calculation  similar to the one in Sec. \ref{sec.atomespres} shows
that ${\cal G}$ has a width
\begin{equation}
\Delta N_i=\frac{1}{\sqrt{\delta \Omega_i \frac{M}{2}g_0^2f^2 N_p}}.
\label{eq.Deltanidens}
\end{equation}

\begin{figure}[ht]
\includegraphics{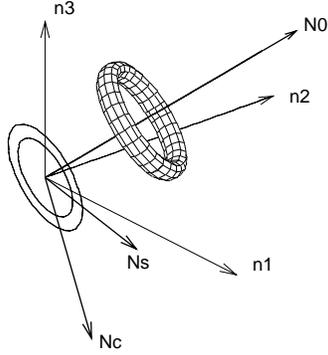}
\caption{\it Distribution over the states
($\ket{n_1,n_2,...,n_M}$) or the states
($\ket{N_0,N_{c_1},N_{s_1},...,N_{c_{(M-1)/2}},N_{s_{(M-1)/2}}}$)
after the measurement. Compare with Fig. 5.
The case of only 3 cells is represented. 
the distribution of the population in the basis
($\ket{N_c,N_s}$) for a well defined $N_0$ does not depend on $N_0$
and is given by the projection of the torus on the plane
$(N_c,N_s)$.}
\label{fig.distrifin}
\end{figure}

Figure \ref{fig.distrifin} depicts the final distribution over
the states $N_0,(N_{c_i},N_{s_i})$ in the simple case
where the cloud is
divided into only three cells.

\subsection{Effect of scattering on $<J_x>$}
Photons scattered in the forward direction in the solid angle
$\theta < \theta_0$ give information on $N_0$ and thus on the total
number of atoms in $\ket{a}$. The other scattered photons give
information on the spatial fluctuations of the atoms in
$\ket{a}$ and, 
in the case we consider where the number of atoms
per cell is large and approximately constant, no information on $N_0$ is
given by the photons scattered outside the solid angle $\theta< \theta_0$.
This is
completely different from the case considered in 
section \ref{sec.atomespres}
where in a single experiment the number of atoms in $\ket{a}$ was
mainly determined by the scattering.

In the case considered here, even if $N_0$ is not affected by the
scattering outside the incident beam, the state is affected by 
scattering events due to the departure from the space of states
which is symmetric under exchange of atoms.  
After a series of detections, the state of the cloud has
components
in different subspaces $J$ because the scattering (or its absence)
brings the system into non symmetric states with $J < N/2$.
We thus expect $\langle J_x \rangle$ to be smaller than the value obtained
for a state  with the same distribution of $J_z$ eigenstates but in the
symmetric subspace $J=N/2$.

We have
\begin{equation}
\langle J_x\rangle
=\sum_i \langle J_{x_i}\rangle
\label{Eq.sumJxi}
\end{equation}
where $\langle J_{x_i}\rangle$ is the mean
value of $J_x$ corresponding to the atoms
of the cell $i$.
It is
\begin{equation}
\langle J_{x_i}\rangle=\int dn_1 ...dn_{i-1}dn_{i+1}...dn_M
\left (\int dn_i |f(n_1,...,n_M)|^2
\right )
\langle J_{x_i}\rangle_{n1,...,n_{i-1},n_{i+1},...,n_M}
\end{equation}
The state is invariant by exchange of atoms inside a single cell,
and the results of Eq.(\ref{gauss}) can be used
to estimate $\langle
J_{x_i}\rangle_{n1,...,n_{i-1},n_{i+1},...,n_M}$ by

\begin{equation}
\langle J_{x_i}\rangle_{n1,...,n_{i-1},n_{i+1},...,n_K}
\simeq \frac{N_{cell}}{2} e^{-\frac{1}{8 \rm{Var}(J_{z_i})}}
\label{Eq.Jxi}
\end{equation}

Taking $i=1$, we note that $\rm{Var}(J_{z_1})$ is the rms width of
the function
\begin{equation}
F(n_1)= {\cal C}'
\left (\frac{n_1}{\sqrt{M}}+ \frac{n_2+ ... + n_M}{\sqrt{M}}\right )
\prod_i {\cal H}_i'\left (\sqrt{N_{c_i}(n_1)^2+N_{s_i}(n_1)^2}\right )
\label{eq.fonctionF}
\end{equation}
where ${\cal C}'(N_0)$ is the final population distribution over the
states $\{\ket{N_0}\}$ and ${\cal H}_i'(N_i)={\cal G}_i(N_i) h^4(N_i)$ is
the final population distribution among the states
$\{\ket{N_{c_i}=\cos(\alpha)N_i,N_{s_i}=\sin(\alpha)N_i}\}$ (It does not
depend on the angle $\alpha$).

For the initial distribution, 
${\cal C}'(n_1/\sqrt{M}+...)$
has  an rms width of $\sqrt{M N_{cell}}/2$ in $n_1$ and
${\cal H}_i\left (\sqrt{N_{c_i}(n_1)^2+N_{s_i}(n_1)^2}\right )=
h^2(\sqrt{2/M}\cos(\vec{K}_i.\vec{r}_1)n_1+...)
h^2(\sqrt{2/M}\sin(\vec{K}_i.\vec{r}_1)n_1+...)$
has an rms width of $\sqrt{M N_{cell}/8}$ as a function of $n_1$.
Thus, Eq.(\ref{eq.fonctionF}) gives an rms width $\Delta J_{z_1}$
equal to $\sqrt{N_{cell}}/2$ for any $(n_2,...,n_M)$, which is
expected.

After the detection, the width of ${\cal C}'$ has been reduced
by the interference measurement to
$\Delta n_{a_{inter}}/\sqrt{M}$
and the rms width in $n_1$ of
${\cal C}'(n_1/\sqrt{M}+...)$ is
\begin{equation}
{\Delta n_1}_{inter}=\Delta n_{a_{inter}}.
\end{equation}
The detection of scattered photons modifies the 
distribution ${\cal H}_i'$  which is then 
no longer separable in $N_{c_i}$ and $N_{s_i}$.
But it follows from the definition (\ref{eq.defNcs}) and from 
Eq.\ref{eq.Deltanidens} that,
in average, ${\cal H}_i'$ has an rms width in $n_1$ which scales as
\begin{equation}
{\Delta n_1}_i \sim \sqrt{M} \Delta N_i
\end{equation}
If we assume that the distribution of $n_1$ is given by the initial Gaussian,
which is multiplied by gaussian factors due to the interferometric
detection and due to the scattering, we obtain the result
\begin{equation}
\frac{1}{\Delta n_1 ^2}=\frac{1}{\Delta n_{a_{inter}} ^2}+\sum_{i,
\vec{K}_i \in S_{scatt}} \frac{1}{M \Delta N_i^2}+
\sum_{i,\vec{K}_i \notin S_{scatt}}\frac{8}{M N_{cell}}
\end{equation}

The squeezing due to scattering and due to interferometric detection
is significant, and we can hence ignore the last term, which is due to 
the initial width of the distribution. 
The contribution due to the interferometric detection is given by
Eq.(18), and the sum over the $i$ for which $\vec{K}_i$ is on the 
scattering sphere follows from Eq.(\ref{eq.Deltanidens}):
\begin{equation}
\sum_{i,
\vec{K}_i \in S_{scatt}} \frac{1}{M \Delta N_i^2} =2\pi g_0^2 f^2 N_p .
\label{just}
\end{equation}
By comparison of Eq.(18) and (\ref{just}), we see that the contribution
due to interferometric detection is approximately a factor $g_0^2/k^2$
times the one due to scattering.
$1/g_0^2$ is larger than  the area of the cloud which itself is 
much larger than $\lambda^2$.  Thus, $g_0^2/k^2 \ll 1$ and
the width in $n_1$ 
is mainly determined by the scattered photons~: 
\begin{equation}
\frac{1}{\Delta n_1^2}
=\frac{1}{\rm{Var}J_{z_1}}\simeq \frac{1}{2}\sigma_1 N_p.
\end{equation}

Coming back to Eq.(\ref{Eq.Jxi}) and Eq.(\ref{Eq.sumJxi}), we get
\begin{equation}
\langle J_x \rangle \simeq
\frac{N}{2}e^{-\frac{\pi}{8}g_0^2 f^2 N_p}
\end{equation}
Thus, the squeezing factor writes
\begin{equation}
\xi=\frac{1}{\sqrt{N N_p} \pi g_0^2 \frac{f}{k}} 
e^{\frac{\pi}{8}g_0^2 f^2 N_p}
\end{equation}
Its minimum value, achieved for $N_p=4/(\pi g_0^2 f^2)$, is 
\begin{equation}
\xi_{Min}=\sqrt{\frac{e}{4\pi N_{at}}}\frac{k}{g_0},
\end{equation}
which is similar to the result obtained for the small cloud and for the
dilute sample.

\section{Conclusion}

We have presented an analysis of the change of the atomic distribution
on internal levels caused by a measurement of the phase shift of an
optical field traversing the atomic sample.
In every run of the detection experiment,
the atomic population statistics is modified in accordance with
the phase shift measurements. 
In practical spin squeezing, the reduced variance in  one of the 
spin components is not the only relevant parameter. one has to 
observe the change of the length of the mean spin $\langle J_x \rangle$ 
as well. The mean spin is reduced, and in case of perfect detection,
we estimate the optimum number of detected photons, and the optimum
value of the squeezing parameter $\xi=\sqrt{1/N_{at}}$, (\ref{xiperf}).

In an experimental implementation, scattering of
photons is inevitable. If these photons are not detected, they will have
no average effect on the atomic populations, but 
scattering leads to further reduction of
the mean spin and an increase of $\xi$.
Two different mechanisms are shown to be responsible for this reduction:
In case of the small cloud, the scattered photons carry information 
on $n_a$, which forces a reduction in $\langle J_x \rangle$. In case of
the dilute cloud, they carry information
on the spatial distribution of the atoms in $\ket{a}$. 
The general problem is difficult to treat, and we focussed on 
three different limiting cases : a cloud of size smaller 
than the wave length, a  dilute cloud where each scattered photon 
can be traced back to a single atom, and a large cloud of density 
larger than $1/\lambda^3$. Both numerical simulations and analytical 
approach were carried out and we showed that scattering decreases
the mean spin $\langle J_x \rangle$. 

Although the physics is very different,
the scattering gives rise to approximately the same optimum
squeezing   
\begin{equation}
\xi_{\rm Min}=\sqrt{1/N_{at}}\,\frac{k}{g_0}.
\label{xifinal}
\end{equation}
which is obtained by detecting the phase shift of a given number of
photons, $N_p\simeq 1/(g_0^2 f^2)$.
We cannot focus to better than within a wavelength, hence the
optimum always exceeds the ideal results (\ref{xiperf}).
In the case of a large cloud, $1/g_0^2$ is of the order of 
the area of the beam at the focus, and $g_0$ 
is limited by $\sqrt{{\cal A}}$, where ${\cal A}$ is 
the area of the cloud . In this case we can therefore rewrite
the expression for 
$\xi_{\rm Min}$ 
\begin{equation}
\xi_{\rm Min} > \sqrt{\frac{\cal A}{N_{at}\lambda^2}}=\sqrt{\frac{1}{D}},
\label{eq.xiOD}
\end{equation}
where $D=N_{at}\lambda^2/{\cal A}$ is the optical density of 
the cloud. Thus for a dense cloud spin squeezing is
possible, whereas for a very dilute cloud we recover the
result that the possibility to know, for any scattered photon, 
which atom it comes, effectively prevents spin squeezing.

 A better squeezing could be achieved if the atomic cloud
lie in an optical cavity so that a photon passes effectively 
$n_t$ times in average through the cloud, where $n_t$ is the
finesse of the cavity.   
 In particular, this would enable squeezing of a dilute cloud.
 To estimate the best possible squeezing in this case, 
two opposit effects should be 
taken into account. 
 First, the width in $n_a$ infered from the phase shift of the
beam after the passage in the cavity is decreased by a factor 
$n_t$ compared to Eq.\ref{eq.Dnaint}.
 Second, the probability that a photon is scattered before
it leaves the cavity is also enhanced by $n_t$ so that the 
number of photons $N_p$ that can be used before $\langle J_x \rangle$ 
decreases too much is also decreased by $n_t$. 
 But, as the width in $n_a$ induced by the phaseshift decreases only 
in $1/\sqrt{N_p}$, the best squeezing factor $\xi$ is still 
decreased by a factor $1/\sqrt{n_t}$ compared to Eq.\ref{xifinal}
and could thus become smaller than 1.


\end{document}